\documentclass[12pt]{iopart}

\usepackage{latexsym, graphicx}
\usepackage[dvips]{color}
\usepackage{iopams}
\usepackage{setstack}

\newcommand{\Realpart}{{\mathrm{\,Re\,}}}

\newcommand{\x}{\mathsf{x}}
\newcommand{\z}{\mathsf{z}}

\def\be{\begin{equation}}
\def\te{\end{equation}}
\def\ee{\end{equation}}
\def\ba{\begin{eqnarray}}
\def\bea{\begin{eqnarray}}

\def\tea{\end{eqnarray}}
\def\ea{\end{eqnarray}}
\def\eea{\end{eqnarray}}

\newskip\humongous \humongous=0pt plus 1000pt minus 1000pt

\newif\ifdtup

\begin{document}

\title{Relativistic Quantum Information in Detectors-Field Interactions}

\author{B. L. Hu$^{1,2}$, Shih-Yuin Lin$^3$ and Jorma Louko$^{4,5}$}

\address{$^1$Maryland Center for Fundamental Physics and Joint Quantum Institute,\\ University of
Maryland, College Park, Maryland 20742-4111 U.S.A.}
\address{$^2$Institute for Advanced Study and Department of Physics,\\
Hong Kong University of Science and Technology, Hong Kong, China}
\address{$^3$Department of Physics, National Changhua University of Education,
Changhua 50007, Taiwan}
\address{$^4$School of Mathematical Sciences, University of Nottingham,
University Park, Nottingham NG7 2RD, United Kingdom}
\address{$^5$Kavli Institute for Theoretical Physics,
University of California, Santa Barbara, CA 93106-4030, USA}
\ead{blhu@umd.edu,sylin@cc.ncue.edu.tw,jorma.louko@nottingham.ac.uk}

\date{\today}

\begin{abstract}
We review Unruh-DeWitt detectors and other models of detector-field interaction in a relativistic quantum field theory 
setting as a tool for extracting detector-detector, field-field and detector-field correlation functions of interest 
in quantum information science, from entanglement dynamics to quantum teleportation. We in particular
highlight the contrast between the results obtained from linear perturbation theory which can be justified provided 
switching effects are properly accounted for, and the nonperturbative effects from available analytic expressions 
which incorporate the backreaction effects of the quantum field on the detector behaviour. 
\end{abstract}




\maketitle

\section{Introduction and Background}
\label{intro}





\subsection{Some Basic Issues in Relativistic Quantum Information}

We assume the readers are somewhat familiar with the Unruh effect \cite{unruh} on the one hand and the basic issues of quantum information on the other \cite{NielsenChuang} and will only highlight the relativistic aspects of both of these topics here. First, the quantum field acting as an environment to the discrete (qubits) or continuous variables (oscillators) in quantum information processing -- we will refer to these point-like physical objects with internal degrees of freedom as detectors (Unruh-DeWitt detector being the familiar one \cite{DeWitt}) -- will necessarily exert environmental influences on the system. Second, the motional states of the detectors (e.g., inertial or  accelerated, uniformly or otherwise) will affect both the quantum decoherence and entanglement dynamics of these detectors.

\subsubsection{Quantum Field Effects}
The presence of a quantum field is unavoidable, as it acts as an ubiquitous environment to the qubits or detectors in question. Two basic issues of quantum information need to be included in one's consideration are:

\noindent{\bf Quantum decoherence}\,\,\,
Coupling to a quantum field can induce decoherence of a single qubit or oscillator, but their  mutual influences mediated by a field  can lessen the degree of decoherence if the two qubits are placed in close range \cite{ASH};\\
\noindent{\bf Entanglement dynamics}\,\,\,
The entanglement between two qubits or oscillators changes in time as their reduced state (after coarse-graining over the field) evolves; it also depends on their spatial separation \cite{ASH,LH09}.

\subsubsection{Kinematical effects}\hspace{.1cm}\\
\noindent{\bf Unruh effect}\,\,\,
A uniformly accelerated detector coupled with a quantum field in the Minkowski vacuum 
would experience a thermal bath of the field quanta at the Unruh temperature proportional to its proper acceleration.
This was first discovered by Unruh using time-dependent perturbation theory \cite{unruh}. Generalized considerations follow in the works of Higuchi et al \cite{Higuchi:1993cya} and Louko et al \cite{satz-louko:curved}.
Exact solutions going beyond these 
test-field descriptions were found by Lin and Hu \cite{LH06,LH07} with interesting new physics.\\
\noindent{\bf Non-uniform acceleration}\,\,\,
The kinematical viewpoint has proven to be more malleable and adaptable than the traditional geometrical (global concepts like event horizon) viewpoint. We will mention how newer models in the 90s such as the RSG model \cite{RSG91}, especially those which incorporate open quantum system concepts such as the RHAK models \cite{RHA,RHK}, have aided in treating non-uniform acceleration, in work from the 90s (e.g., \cite{Svaiter:1992xt}) to now \cite{OLMH11}.\\
\noindent{\bf Mutual influences}\,\,\,
The influence of one detector on the field will propagate in space and affect other detectors after
some time. 
These causal mutual influences propagating back and forth is a source of non-Markovianity in multi-detector theories.
They act to augment the quantum coherence between two detectors placed in close range.
Another source of non-Markovianity is the long range autocorrelation of the quantum field. 

\subsubsection{Relativistic Effects}

Furthermore, objects in a relativistic system may behave differently when observed in different reference frames,
so we have:\\
\noindent{\bf Frame dependence}\,\,\,
Quantum entanglement of two objects localized at different positions on a spatial hypersurface is a kind of 
spacelike correlation, the time evolution of their entanglement will depend on how the spacetime is foliated by 
spacelike hypersurfaces.\\
\noindent{\bf Time dilation}\,\,\,
For moving objects with worldlines parameterized by their proper times, 
their time dilations observed in a reference frame will naturally enter the dynamics in that frame.\\
\noindent{\bf Projective measurement local in space}\,\,\,
Quantum states make sense only in a given frame where a Hamiltonian is well defined \cite{AA81}.
Two quantum states of the same system with quantum fields in different frames are comparable only on those totally 
overlapping time-slices associated with certain moments in each frame.
By a measurement local in space, e.g. on a point-like UD detector coupled with a quantum field,
quantum states of the combined system in different frames can be interpreted as if they collapsed on different
time-slices passing through the same measurement event. Nevertheless, the post-measurement states will evolve to
the same state up to a coordinate transformation when they are compared at some time-slice in the future. 
In a two-detector system with the first detector being measured at some moment, the reduced state 
of the second detector collapsed in different frames will become consistent once it enters 
the future lightcone of the measurement event.
\cite{ASH, LH09, LCH08, Lin11a, LSCH12}.

\subsection{Unruh effect via perturbation theory and quantum information via exact solutions}

Time-dependent perturbation theory (TDPT) was used by Unruh originally to show the detector response to uniform acceleration. For a comprehensive description of Unruh effect, see, e.g., the recent review of \cite{Crispino:2007eb}.
With the infusion of quantum open system ideas in the 90s these TDPT results were later found to hold only in the Markovian regime, 
corresponding to the limits of ultra-high acceleration or ultra-weak coupling. Discovery of exact solutions in the 2000s showed 
that the transition probability calculated from the infinite-time TDPT is valid only in transient under restricted conditions.
We will develop the perturbative theory further in Section 2 and comment on these developments in the last two sections.
In Section 3 we will introduce two other more general models for moving detector- quantum field interaction, namely, the RSG and the RHAK models,
treating the detectors as harmonic oscillators rather than the two level system as in the original Unruh derivation. 
We will also bring in the  broader scope provided by the theory of open quantum systems exemplified by the quantum Brownian model, 
where the use of reduced density matrix and influence functionals opens the way to exploring the full parameter range of detector-field 
interaction including self-consistent backreaction. This opens the door for quantum information inquires. 
In Section 4 we describe the detector-field dynamics from the exact solutions of one such model, 
and identify the limitations of TDPT.  In Section 5 we give an important example of relativistic quantum information, 
that of quantum teleportation which uses pretty much all of the relativistic and quantum information elements developed, 
such as frame dependence and entanglement dynamics.  We end with some suggestions on further developments.


\section{Nonstationary detector within first-order perturbation theory}
\label{sec:scalar-pointlike-linear}

In this section we summarise recent results about the
transition rate of a pointlike
detector within linear perturbation theory,
in situations where neither the detector trajectory nor
the state of the quantum field is assumed 
stationary~\cite{satz-louko:curved,satz:smooth,Hodgkinson:2011pc,hodgkinson-louko-btz}.
The central issue is to isolate switch-on and switch-off effects
from phenomena that are genuinely due to the
acceleration and to the state of the field.

\subsection{Transition probability}
\label{subsec:1storderptheory}

We consider a pointlike {\it two-level} detector that moves in a spacetime of dimension $d\ge2$ along the 
worldline~$\z(\tau)$, where the parameter $\tau$ is the detector's proper time. The motion is driven by an external 
agent who is decoupled from the detector's internal degrees of freedom and from the quantum field to which the detector couples.

The detector's internal Hilbert space here is two-dimensional, spanned by the orthonormal basis states
$|0\rangle_d$ and $|\omega\rangle_d$ whose respective energy eigenvalues are $0$ and~$\omega$, with $\omega\ne0$. 
For $\omega>0$, $|0\rangle_d$ is the ground state and $|\omega\rangle_d$ is the excited state; for $\omega<0$, 
the roles of the states are reversed. A~generalisation to a countable number of nondegenerate energy eigenstates would be straightforward.

The spacetime contains a free real scalar field~$\phi$,
whose mass and
curvature coupling parameter may be arbitrary.
The detector is coupled to $\phi$ linearly,
by the interaction picture Hamiltonian
\begin{equation}
H_{{\mathrm{int}}}=\lambda\chi(\tau)Q(\tau)\phi\bigl(\z(\tau)\bigr)
\ ,
\end{equation}
where $\lambda$ is the coupling constant and $Q$ is
the detector's monopole moment operator. The switching function $\chi$ specifies how the interaction is turned on and off. We assume $\chi$ to be smooth, nonnegative and of compact support. We also assume the trajectory
$\z(\tau)$ to be smooth.

We denote the initial state of the field by $|\psi_0\rangle$,
and we assume $|\psi_0\rangle$ to be
regular in the sense of the Hadamard property~\cite{kay-wald,Decanini:2005gt}.
The detector is initially prepared in the state~$|0\rangle_d$.

We work within first-order perturbation theory in~$\lambda$.
After the interaction has ceased, the probability for the detector to be found in the state $|\omega\rangle_d$,
regardless the final state of the field, is \cite{byd,wald-smallbook}
\begin{equation}
\label{eq:prob}
P(\omega)=\lambda^2{|_d\langle0|Q(0)|\omega\rangle_d|}^2\mathcal{F}\left(\omega\right)
\ ,
\end{equation}
where the response function $\mathcal{F}\left(\omega\right)$
is given by
\begin{equation}
\label{eq:respfunc-alt}
\mathcal{F}\left(\omega\right)=2 \Realpart
\int^{\infty}_{-\infty}\,\mathrm{d}u\,\chi(u)\int^{\infty}_0\,\mathrm{d}s\,\chi(u-s)\,e^{-i\omega s} \, W(u,u-s)
\ ,
\end{equation}
and the correlation function
$W(\tau',\tau'') :=\langle\psi_0|\phi\bigl(\z(\tau')\bigr)\phi\bigl(\z(\tau'')\bigr)|\psi_0\rangle$ is the pull-back of the Wightman function
to the detector's worldline.
The prefactor $\lambda^2{|_d\langle0|Q(0)|\omega\rangle_d|}^2$ in
(\ref{eq:prob})
depends only on the detector's
internal structure, while
$\mathcal{F}\left(\omega\right)$
(\ref{eq:respfunc-alt}) encodes the dependence on~$|\psi_0\rangle$,
the trajectory and the switching. With minor abuse of terminology,
we refer to $\mathcal{F}\left(\omega\right)$ as the transition probability.

\subsection{Transition probability without distributional integrals}
\label{subsec:regulator-free}

While formula (\ref{eq:respfunc-alt}) for the transition probability
is as such well defined, it is not well suited for discussing
how the probability depends on the switching function,
especially when the switching becomes sharp.
The correlation function $W$ is not a genuine function but a distribution.
When $W$ is represented by a family
$W_\epsilon$ of functions that converge to
$W$ as $\epsilon\to0_+$,
the sense of convergence entails that the limit
$\epsilon\to0_+$ is taken in (\ref{eq:respfunc-alt})
only \emph{after\/} the integrals are
evaluated~\cite{hormander-vol1,hormander-paper1,Fewster:1999gj, junker}.
The sharp switching limit may hence not necessarily be brought under the
integrals and the
$\epsilon\to0_+$ limit in
(\ref{eq:respfunc-alt})~\cite{schlicht,Schlicht:thesis,louko-satz:profile}.

What is needed is to re-express (\ref{eq:respfunc-alt}) in terms of the
genuine function $W_0 := \lim_{\epsilon\to0_+}W_\epsilon$, where the limit
is understood pointwise.
The results for $d=2$, $d=3$ and $d=4$ are 
\cite{satz-louko:curved,satz:smooth,Hodgkinson:2011pc,hodgkinson-louko-btz}
\begin{eqnarray}
\fl
\label{eq:2respfunc}
\mathcal{F}_{d=2}\left(\omega\right)
&=
2 \Realpart
\int^{\infty}_{-\infty}\,\mathrm{d}u\,\chi(u)\int^{\infty}_0\,\mathrm{d}s\,\chi(u-s)\,e^{-i\omega s} \, W_0(u,u-s)
\ ,
\\[1ex]
\fl
\mathcal{F}_{d=3}(\omega)
&=
\frac{1}{4}\int^{\infty}_{-\infty}\,\mathrm{d}u\,\left[\chi(u)\right]^2
+2\int_{-\infty}^{\infty}\mathrm{d}u\,\chi(u)\,\int_{0}^{\infty}\mathrm{d}s\,\chi(u-s) \Realpart \! \left[\mathrm{e}^{-i\omega s}W_0(u,u-s)\right]
\ ,
\nonumber
\\
\fl
\label{eq:resp-had-comp}
\\
\fl
\mathcal{F}_{d=4}(\omega)
&=
-\frac{\omega}{4\pi}\int_{-\infty}^{\infty}\mathrm{d}u\,{[\chi(u)]}^2
\ + \
\frac{1}{2\pi^2}\int_0^{\infty}
\frac{\mathrm{d}s}{s^2}\int_{-\infty}^{\infty}\mathrm{d}u\,\chi(u)
\bigl[ \chi(u)-\chi(u-s)\bigr]
\nonumber
\\[1ex]
\fl
&
\hspace{3ex}
+
2
\int_{-\infty}^{\infty}\mathrm{d}u\,\chi(u)
\int_0^{\infty}\mathrm{d}s\,\chi(u-s)
\Realpart \! \left(
\mathrm{e}^{-i\omega s}\, W_0(u,u-s)+\frac{1}{4\pi^2s^2}
\right)
\,,
\label{probability}
\end{eqnarray}
and those for
$d=5$ and $d=6$ can be found in \cite{hodgkinson-louko-btz}
in the special case of a Minkowski space massless field in the Minkowski vacuum.
The crucial point is that in addition to an expected integral term that involves~$W_0$,
there are also additional terms that depend on the switching.
These additional terms are remnants of the distributional singularity of~$W$,
and they are absent only for $d=2$,
where the singularity of $W$ is merely logarithmic.

The Hadamard property of the Wightman function guarantees that
the integrals in
(\ref{eq:2respfunc})--(\ref{probability})
are convergent at $s=0$.
We assume that any singularities of $W_0$ at $s>0$ are integrable.
Such singularities can occur for example when the spacetime
has spatial periodicity so that points on the detector's
trajectory can be joined by null geodesics that
circumnavigate the space~\cite{hodgkinson-louko-btz}.

\subsection{Transition rate}
\label{subsec:sharp}

When both the detector trajectory and the quantum state of the field
are stationary, in the sense that they are invariant under a Killing
vector that is timelike in a neighbourhood of the trajectory, a
transition rate per unit time may be defined by making the
switching function time-independent and
formally factoring out the infinite total time of detection
\cite{unruh,DeWitt,Crispino:2007eb,byd,wald-smallbook,Hartle:1976tp,gibb-haw:dS,Letaw:1979wy,Letaw:1980yv,takagi,Sonego:2003zh}.
In time-dependent situations this procedure is however not available,
and separating the switching effects from the acceleration effects
becomes delicate
\cite{Higuchi:1993cya,Svaiter:1992xt,Letaw:1980yv,Hinton:1984ht,grove:1988-add,Sriramkumar:1994pb,Suzuki:1997cz,cande-sciama,davies-ottewill}.

To define a transition rate in the nonstationary setting,
we consider the limit in which the detector is switched on an off sharply.
We let the switching function $\chi$ take the value unity
from proper time $\tau_0$ to proper time~$\tau$, where $\tau_0<\tau$,
and we assume that the switch-on takes
place over an interval of duration $\delta$
before~$\tau_0$ and the
switch-off takes
place over an interval of duration $\delta$
after~$\tau$, in a manner
discussed in~\cite{satz-louko:curved,satz:smooth}.
The limit of sharp switching is $\delta\to0$.

We regard the response function $\mathcal{F}$
as a function of the switch-off moment~$\tau$,
and we define
$\dot{\mathcal{F}}_{\tau} :=
\mathrm{d}\mathcal{F}/\mathrm{d}\tau$.
$\dot{\mathcal{F}}_{\tau}$ may be regarded as the detector's
instantaneous transition rate per unit proper time, observationally
meaningful in terms of consequent measurements in identical
ensembles of detectors~\cite{satz-louko:curved}.


For $d=2$ and $d=3$, taking the the $\delta\to0$ limit in
(\ref{eq:2respfunc}) and (\ref{eq:resp-had-comp}) is
immediate and yields a finite result for the transition probability.
For $d=4$, the $\delta\to0$ limit in (\ref{probability})
contains a divergent term proportional to $\ln\delta$~\cite{satz-louko:curved,satz:smooth}.
This divergent term depends on the details of the switching but it is constant in time,
and it is also independent of the trajectory and of the quantum state.
The divergent term does hence not contribute to the transition rate.
Physically, the $\delta\to0$ limit means that we take the switching to
be rapid compared with the overall duration of the interaction:
focusing on the transition rate allows us to discard from the
transition probability the numerically dominant piece that only
depends on the details of the switching.
Collecting, the $\delta\to0$ transition rates for $d=2$, and $d=3$ and $d=4$
are given by
\begin{eqnarray}
\fl
d=2: \hspace{2ex}
\dot{\mathcal{F}}_{\tau}\left(\omega\right)=2 \Realpart
\int^{\Delta\tau}_0\,\mathrm{d}s\,\,e^{-i\omega s} \, W_0(\tau,\tau-s)
\ ,
\label{eq:2tranrate}
\\[1ex]
\fl
d=3: \hspace{2ex}
\dot{\mathcal{F}}_{\tau}\left(\omega\right)
=
\frac{1}{4}+2\int^{\Delta\tau}_{0}\,\mathrm{d}s\,
\Realpart\left[\mathrm{e}^{-i\omega s}W_0(\tau,\tau-s)\right]
\ ,
\label{eq:3tranrate}
\\[1ex]
\fl
d=4: \hspace{2ex}
\dot{\mathcal{F}}_{\tau}(\omega)
=
-\frac{\omega}{4\pi}
+ 2\int_0^{\Delta\tau}\mathrm{d}s
\Realpart \!
\left( \mathrm{e}^{-i\omega s}W_0(\tau,\tau-s)+\frac{1}{4\pi^2s^2}\right)
\ +\frac{1}{2\pi^2 \Delta \tau}
\ ,
\label{eq:4tranrate}
\end{eqnarray}
where $\Delta\tau:=\tau-\tau_0$.
(\ref{eq:2tranrate})~and (\ref{eq:3tranrate}) are valid as $\delta\to0$ at fixed~$\lambda$,
provided $\lambda$ is so small that the total transition probability remains within the validity domain of the perturbative treatment.
(\ref{eq:4tranrate})~is valid as $\delta\to0$ provided $\lambda$ simultaneously approaches zero
so fast that it is bounded in absolute value by $k/\sqrt{|\ln\delta|}$,
where the positive constant $k$ is so small that the total transition probability remains within the validity domain of the perturbative treatment.

For $d=5$ and $d=6$, we
specialise to a massless field in
Minkowski spacetime
in the Minkowski vacuum~\cite{Hodgkinson:2011pc}.
The transition probability
contains again a term that diverges as $\delta\to0$.
For $d=5$ the divergent term is constant in time,
and the transition rate has the finite $\delta\to0$ limit
\begin{eqnarray}
\fl
d=5: \hspace{2ex}
\dot{\mathcal{F}}_{\tau}\left(\omega\right)
=
\frac{4\omega ^2 + \ddot{\z}^2(\tau)}{64\pi}
\ + \
\frac{1}{4\pi^2}\int^{\Delta\tau}_{0}\,\mathrm{d}s
\left(\frac{\sin{(\omega s)}}{\sqrt{\left[-{(\Delta\z)}^2\right]^3}}-\frac{\omega}{s^2}\right)
\ -\frac{\omega}{4\pi^2\Delta\tau}
\ ,
\nonumber
\\
\fl
\phantom{x}
\end{eqnarray}
where $\Delta\z := \z(u) - \z(u-s)$.
For $d=6$, by contrast, even the
transition rate contains a term that diverges for generic trajectories
as $\delta\to0$,
proportionally to $\ddot{\z}\cdot \dddot{\z} \ln\delta$.
This means that the divergences due to the rapid switching cannot be
isolated from the acceleration effects for $d=6$.
The sole exception occurs for
trajectories whose scalar proper acceleration $\sqrt{\ddot{\z}^2}$
is a constant, including as a special case all stationary trajectories.
For such trajectories the $d=6$ transition rate remains
finite as $\delta\to0$ and is given by
\begin{eqnarray}
\fl
d=6: \hspace{2ex}
\dot{\mathcal{F}}_{\tau}(\omega)
&=
-\frac{\omega\bigl(\omega^2+\ddot{\z}^2\bigr)}{24\pi^2}
\ + \
\frac{1}{2\pi^3}\int^{\Delta\tau}_{0}\,\mathrm{d}s\,\left(
\frac{\cos{(\omega s)}}{\left[(\Delta \z)^2\right]^2}
-\frac{1}{s^4}+\frac{3\omega^2+\ddot{\z}^2}{6s^2}\right)
\nonumber
\\[1ex]
\fl
&\hspace{3ex}
+\frac{3\omega^2+\ddot{\z}^2}{12\pi^3\Delta\tau}
\ - \
\frac{1}{6\pi^3\Delta\tau^3}
\ .
\label{eq:ss:6d:transrate}
\end{eqnarray}

\subsection{Applications}

When both the detector trajectory and the quantum state of the field are stationary,
the transition rate formulas (\ref{eq:2tranrate})--(\ref{eq:ss:6d:transrate})
reduce to the well-known formulas
in which stationarity is assumed at the outset~\cite{byd,wald-smallbook,takagi}.
We re-emphasise, however, that formulas
(\ref{eq:2tranrate})--(\ref{eq:ss:6d:transrate})
apply in genuinely time-dependent situations.

A showcase example is a Minkowski spacetime trajectory that is
asymptotically inertial at early times and of asymptotically uniform
linear acceleration at late times, with the field in the Minkowski vacuum. Within the perturbative treatment, the transition rate is duly found \cite{louko-satz:profile}
to interpolate between that in inertial motion and that in uniform linear acceleration, describing thus the onset of the Unruh effect~\cite{unruh}.

Other applications can be found
in~\cite{satz-louko:curved,satz:smooth,hodgkinson-louko-btz,louko-satz:profile,Dragan:2011zz}.
Slowly-varying acceleration
is discussed in~\cite{Obadia:2007qf,Kothawala:2009aj}.

\subsection{Other definitions of the transition rate}

To end this section, we mention two alternative definitions of the
transition rate.

First, the transition
rate of a pointlike detector in flat spacetime can be defined by
first giving the detector a spatial size, specified
covariantly in terms of the detector's instantaneous rest frame,
and at the end taking the pointlike limit~\cite{schlicht,Schlicht:thesis,takagi}.
The results agree with those obtained via smooth switching
in the common domain of validity
\cite{Hodgkinson:2011pc,schlicht,Schlicht:thesis,louko-satz:profile,Langlois,Langlois-thesis}.
A~related procedure that replaces spatial size by a
pole prescription in proper time is discussed in~\cite{Obadia:2007qf}.

Second, a spatially extended detector in flat spacetime
can be reinterpreted as a pointlike detector with an
energy cutoff that is specified covariantly in the
detector's instantaneous rest frame~\cite{Langlois,Langlois-thesis}.
Definition of the transition rate via this energy cutoff can be generalised to curved spacetimes at least when the spacetime has a sufficient
amount of symmetry~\cite{Langlois,Langlois-thesis}.

\section{Detector-Field Interaction: UD, RSG, RHAK models}


Quantum mechanics for the single particles in some non-linear potentials such as anharmonic oscillators,
Morse potential, etc. are exactly solvable. But in field theory, since a field has infinitely many degrees of
freedom, a small non-linearity can create huge difficulty in calculations. One can at most do
perturbation theory or self-consistent approximations around some non-trivial background field configuration,
where the calculation involves essentially Gaussian integrals.
However, if the potential of the detector is that of a harmonic oscillator (HO) and the quantum state is in a Gaussian form,
it is possible to solve the full dynamics of the combined system of the detectors and the field  non-perturbatively.


\subsection{Raine-Sciama-Grove (RSG) Model}

A notable non-perturbative model  is that of Raine, Sciama, and Grove \cite{RSG91}, where they found the late-time expectation values for the stress tensor of a massless scalar field in (1+1)D Minkowski space.
The method is generalized to the case with a point-like UD detector in a massless scalar field
in (3+1)D Minkowski space, and the whole history of the combined system is solved in \cite{LH06}.

\subsection{Proxy to Quantum Brownian Motion Models}

Suppose the internal degrees of freedom of the UD detector are HOs.
Then the combined system of $N$ UD detectors and a quantum field is an $(N + \infty)$-HO system, which is linear and exactly solvable. Unruh and Zurek \cite{UZ89} have studied a model where a hamonic oscillator interacts with a massless scalar field in 2-D. They derived the exact master equation for the reduced density matrix of the system (oscillator) at a temperature determined by the initial state of the field, and observed some
general features different from the conventional Markovian results  valid for an ohmic bath at ultra-high temperature  made known earlier in the famous paper of  Caldeira and Leggett \cite{CalLeg83}. One feature is the dependence of the ultraviolet cut-off in the master equation and the reduced density matrix, and thus also in the von Neumann entropy of the system.

More general non-Markovian behavior was explored by Hu, Paz and Zhang \cite{HPZ} who derived an exact master equation with nonlocal dissipation and colored noise for the system of one harmonic oscillator (detector) interacting with a thermal bath of n-harmonic oscillators. For quantum decoherence they identified the low temperature, supra-ohmic regime as a noticeable departure from the Markovian behavior (see followup in \cite{PHZ93}).
Using this model as a theoretical tool with the help of quantum open system ideas,
many of the basic issues we listed in the beginning can be addressed effectively.

Generalizing a bath of n-harmonic oscillators of time-dependent frequencies to a quantum field was subsequently done by 
Hu and Matacz \cite{HM} for moving detectors- quantum field interactions. 
Because the treatment is given in quantum optics language -- the quantum states described by the squeeze, rotation and 
displacement operators and the dynamics in terms of parametric amplification -- their results are immediately applicable to 
``atomic-mirror"-optical systems \cite{GBH12}.
It also served the intended purpose of bringing open systems methods and concepts to quantum field theory.
The influence functional treatment they used incorporates the backreaction of
the environment on the system (which could be either the quantum
field or the harmonic oscillator depending on what one is after)
in a self-consistent way. In particular they showed how the Unruh and Hawking temperatures  can be identified from the noise kernel using this method.

Viewing the Unruh effect from this perspective, since in the QBM model there are nontrivial activities at zero temperature \cite{UZ89,HPZ,HM}, 
we note that even for the zero acceleration $a=0$ case the detector is not just laying idle but has interesting physical features due to its
interaction with the vacuum fluctuations in the quantum field.

\subsection{Raval-Hu-Anglin-Koks (RHAK) Models}

A model of N detectors in arbitrary relativistic motion interacting with a common quantum field (but not with each other)
was proposed by Raval, Hu, Anglin \cite{RHA}. They calculated the influence of quantum fields on  the detectors in motion,
and the mutual influence of detectors by the action of fields via the Langevin equations derived from the influence functional.
They introduced the notion of self and mutual impedance, advanced and retarded noise, and the new relations between noise-correlations and dissipation-propagation. They show the existence of general fluctuation-dissipation relations, and for trajectories without event horizons, correlation-propagation relations.  Raval, Hu and Koks \cite{RHK} used this model to explore different trajectories of the moving detectors in a quantum field and showed that this is a more feasible way (over the traditional global  geometric view which relies on the existence of event horizons) to address situations where the spacetime possesses an event horizon only asymptotically, or none at all. Examples studied there include detectors moving at uniform acceleration only asymptotically or for a finite time, a moving mirror, and a two-dimensional collapsing mass. They show that in such systems radiance indeed is observed, albeit not in a precise Planckian spectrum. The setups in this model have been adopted in the study of charge particle motion by Johnson et al \cite{JH1} in an electromagnetic field and by Galley et al \cite{GH} for the self-force of masses moving in a gravitational field.

\subsection{Moving Detectors-Quantum Field Interaction}
\label{DFInt}

Consider a model with N identical point-like Unruh-DeWitt detectors with the internal degrees of freedom
represented by harmonic oscillators with mass $m_0$ and natural frequency $\Omega$, moving in a quantum field
in (3+1)D Minkowski space.  Here we follow the treatment in \cite{LCH08}.
The action of the combined system is given by
\begin{eqnarray}
  S &=& -\int d^4 x \sqrt{-g} {1\over 2}\partial_\sigma\Phi(\x) \partial^\sigma\Phi(\x) +
    \sum_{{\bf d}}\int d\tau_{\bf d} \left\{ {m_0\over 2}\left[\left(\partial_{\bf d}Q_{\bf d}\right)^2
    -\Omega_{0}^2 Q_{\bf d}^2\right] \right. \nonumber\\ & & \hspace{1cm}\left. +\lambda\int d^4 x
    Q_{\bf d}(\tau_{\bf d})\Phi (\x)\delta^4\left(\x-\z_{\bf d}^{}(\tau_{\bf d})\right)\right\},
  \label{Stot1}
\end{eqnarray}
where $\sigma=0,1,2,3$, $g_{\sigma\sigma'} = {\rm diag}(-1,1,1,1)$, ${\bf d}=A, B, C,\cdots $ denotes the
names of the detectors, $\partial^{}_{\bf d}\equiv \partial/\partial \tau^{}_{\bf d}$,
$\tau^{}_{\bf d}$ is the proper time for detector $Q_{\bf d}$ and $\z_{\bf d}^{}(\tau_{\bf d})$ is the
trajectory of detector {\bf d}. The scalar field $\Phi$ is assumed to be massless, and $\lambda$ is the
coupling constant. We consider a massless scalar field here because it is simpler and a good
representation of the electromagnetic field. In fact all kinds of fields, massless or massive,
bosonic or fermionic, can be considered, depending on the physics one aims at.
The detectors do not have to be uniformly accelerated (e.g., \cite{RHK}) or at rest. However,
the motion of the detector here are assumed to be controlled by external agents, in other words,
the trajectories or worldlines of the detectors are prescribed and not dynamical.
If the motion of the detector becomes dynamical, it is extremely hard to get analytical results
even in classical theory (for example, a relativistic charge in a field in classical \cite{Ro65}
and quantum field theory \cite{IYZ11}), since including backreaction of the field on the detector will
alter its (test-field) prescribed trajectory. Trajectories of charged particles \cite{JH1} and
even extended objects \cite{GHL} determined by their interplay with the quantum field have also
been studied before using the influence functional method which is particularly suited to
treating consistent backreaction effects.

The first noticeable attractive feature of (\ref{Stot1})
is that this model is linear and thus easy to treat.
It is arguably the simplest model for an ``atom"-field interacting system but complex enough to give nontrivial results and
insights. By ``atom" here we refer  to a spatially-localized physical object with internal degrees of freedom.
By field, we categorically refer to dynamical variables which can be non-local in space
\footnote{`Nonlocal' is in the sense used by the atomic-optical quantum information community.
Of course quantum field theory is local.}.
In some simple setups  analytic results can be obtained in the whole parameter range,
and back-reaction to both the atom and the field can be fully studied with the help of quantum open systems techniques.

Moreover, in a relativistic setting, such as for uniformly accelerated detectors or black holes, event horizons for the detectors can be sharply defined since the detectors are always localized. 
Also since the detectors are pointlike, they are allowed to be parametrized by their own proper times, which are
invariants under coordinate transformations. This greatly simplifies the calculations in different reference frames
when the related physics correspond only to the two-point correlators of the detectors parametrized by two
proper times. Note that this is even plausible for extended objects in the spirit of effective field theory \cite{GHL,GHeft}.

For a uniformly accelerated UD detector in (3+1)D with proper
acceleration $a$ the Unruh effect \cite{unruh, DeWitt, Crispino:2007eb, byd} attests
that it should behave the same way as an inertial UD detector in
contact with a thermal bath at Unruh temperature $T_U$, or more
precisely, as an inertial harmonic oscillator in contact with an
Ohmic bath at $T_U$ \cite{FHR}.
However, examining this from the vantage point of the exact solutions
we obtained,  we see the above statements are accurate only at the
initial moment. After the coupling is switched on, the quantum state
of the field will have been changed by the detector, so the field is
no longer in the Minkowski vacuum and it does not make exact sense to
say that the detector is immersed in a thermal state (or any state
defined in the test-field description, i.e., where the field is
assumed not to be modified by the presence of the detector).

A theorem by Bisognano and Wichmann (BW) \cite{BW75} states that the
Minkowski vacuum, which is uniquely characterized by its invariance
under all Poincar\'{e} translations, is a Kubo-Martin-Schwinger (KMS)
state with respect to all observables confined to a Rindler wedge.
It does not apply here because 
the BW theorem refers to the vacuum state of a quantum field alone, not the
combined detector-field system. Even when the combined system is in
a steady state, the quantum state of the interacting field is not
invariant under spatial translations in Minkowski space, hence
does not subscribe to the assumption of the BW theorem pertaining to
Poincar\'{e} invariance.
Actually the Planck factor in, for example, (Eq.(60) in Ref.\cite{LH06}),
\begin{equation}
  \langle Q(\eta)Q(\eta')\rangle_{\rm v} \sim
  {\lambda^2 \hbar\over (2\pi)^2 m_0^2 } \int
  {\kappa d\kappa\over 1-e^{-2\pi \kappa/a}}[\ldots], \label{planck}
\end{equation}
is a consequence of the BW theorem. Nevertheless, it is derived from only the
free-field-solution part of the complete interacting field.
Here the factor is not distorted by the interaction simply because the field is linear and
the coupling is bilinear. For nonlinear fields or couplings it would
have a nonPlanckian spectrum and the departure from the conventional
picture would be more pronounced. 


\section{Nonperturbative Detector-Field Dynamics}

Nonperturbatively solvable models such as (\ref{Stot1}) are particularly useful for examining the full features
of a system which perturbative theories miss or misrepresent.
They are essential for understanding new
physics such as that associated with quantum entanglement whose dynamical behavior we don't
really have a complete or accurate knowledge about.
We now continue to develop the model (\ref{Stot1}). 

A quantum state of the combined detector-field system can be described by the density matrix
$\bar{\rho}[({\bf Q},\Phi_{\bf x}), ({\bf Q}',\Phi'_{\bf x});x^0]$  or equivalently, 
the Wigner function \cite{UZ89, HPZ}
\footnote{Here we write $({\bf Q},\Phi_{\bf x})={\bf \Sigma}-({\bf \Delta}/2)$ and 
$({\bf Q}',\Phi'_{\bf x})= {\bf \Sigma}+ ({\bf \Delta}/2)$ with the boldface letters ${\bf \Sigma}$ 
and ${\bf \Delta}$ denoting the vectors in the configuration space. $\int {\cal D}{\bf \Delta}$ and
$\int {\cal D}{\bf \Sigma}$ are functional integrals.},
\begin{equation}
  W[{\bf P}, {\bf \Sigma}; x^0] = \int {\cal D}\left({\bf \Delta}\over 2\pi\right) e^{{i\over\hbar} {\bf P}\cdot {\bf \Delta}}
    \bar{\rho}\left[ {\bf \Sigma} - {{\bf \Delta}\over 2}, {\bf \Sigma} + {{\bf \Delta}\over 2} ; x^0 \right]
\end{equation}
If we start with a Gaussian state, by virtue of the linearity of the combined system (\ref{Stot1}), the quantum state will always
evolve in a Gaussian form in its entire history. Thus solving the dynamical equations for the Wigner function boils down
to solving the time-dependent factors in the Wigner function. 

Since the field variables at some moment $x^0$ are defined on the whole time-slice associated with $x^0$,
the density matrix or Wigner functions at $x^0$ is also defined on that time-slice.
In the Schr\"odinger picture, the evolution of a 
density matrix (Wigner function) is governed by the
master equation (Fokker-Planck equation). It is possible to solve these equations for Gaussian states directly
in simple cases (e.g. \cite{FRH11,FH12}). However, when the degrees of freedom of the density matrix is large
or even infinite, it becomes very difficult to solve the coupled equations, as the dynamics is often non-Markovian
(the master equations for the reduced state) or non-linear in appearance (the dynamical equations for the
time-dependent factors in the Wigner functions).
Moreover, even the solutions are obtained
and can be expressed formally, the factors in the Wigner function are inverse matrices with infinite dimension,
which are computationally challenging.

To get rid of these difficulties it is convenient to apply the $(K,\Delta)$-representation \cite{UZ89}                 
(or called the Wigner characteristic function \cite{GZ99}), which is a double-Fourier-transformed function of
(thus equivalent to) the usual Wigner function,
\begin{eqnarray}
  & &\rho[{\bf K}, {\bf \Delta}; x^0] = \int {\cal D}{\bf \Sigma}\, e^{{i\over\hbar} {\bf K}\cdot {\bf \Sigma}}
    \bar{\rho}\left[ {\bf \Sigma} - {{\bf \Delta}\over 2}, {\bf \Sigma} + {{\bf \Delta}\over 2} ; x^0 \right]
    =\exp \left[ {i\over\hbar}\left(\langle\hat{\Phi}^{}_\mu\rangle K^\mu
    \right.\right. \nonumber\\ && \hspace{1cm}\left.\left.
    -\langle\hat{\Pi}^{}_\mu\rangle \Delta^\mu\right) -{1\over 2\hbar^2} \left( K^\mu {\cal Q}_{\mu\nu} K^\nu
    -2 \Delta^\mu {\cal R}_{\mu\nu} K^\nu + \Delta^\mu {\cal P}_{\mu\nu} \Delta^\nu\right)
    \right],
\label{Qstate}
\end{eqnarray}
where we denote $\hat{Q}_{\bf d}$ and $\hat{P}_{\bf d}$ by $\hat{\Phi}^{}_{\bf d}$ and $\hat{\Pi}^{}_{\bf d}$, respectively
($\hat{P}_{\bf d}$, $\hat{\Pi}^{}_{\bf x}$ are conjugate momenta to $\hat{Q}_{\bf d}$, $\hat{\Phi}^{}_{\bf x}$),
$\mu, \nu = \{{\bf d}\}\cup\{ {\bf x}\}$ run over all the detector- and field- degrees of freedom defined on the whole time-slice,
and the time-dependent factors ${\cal Q}_{\mu\nu}(x^0)$, ${\cal P}_{\mu\nu}(x^0)$, and ${\cal R}_{\mu\nu}(x^0)$ are exactly the
symmetrized two-point correlators $\langle A,B\rangle \equiv \langle AB+BA \rangle/2$ of the
dynamical variables evaluated on the $x^0$-slice, for they are obtained by, e.g.,
\begin{eqnarray}
  \langle \delta\hat{\Pi}^{}_\mu (x^0), \delta\hat{\Phi}^{}_\nu (x^0)\rangle &=&
    \left. {i\hbar\delta\over \delta \Delta^\mu}{\hbar\delta\over i\delta K^\nu}
    \rho[{\bf K}, {\bf \Delta};x^0] \right|_{{\bf \Delta} = {\bf K}=0} = {\cal R}_{\mu\nu},
\end{eqnarray}
where $\delta\hat{\Phi}^{}_\mu \equiv \hat{\Phi}^{}_\mu - \langle\hat{\Phi}^{}_\mu\rangle$, and 
$\delta\hat{\Pi}^{}_\mu \equiv \hat{\Pi}^{}_\mu -\langle\hat{\Pi}^{}_\mu\rangle$.
Note that in (\ref{Qstate}), ${\cal Q}_{\mu\nu}$ and ${\cal P}_{\mu\nu}$ are defined as symmetric matrices,
but ${\cal R}_{\mu\nu}\not= {\cal R}_{\nu\mu}$ in general.

The reduced states of the system with the environment integrated out are simple for Gaussian states
in the $(K,\Delta)$-representation since a Gaussian integral gives another Gaussian function.
For example, the reduced state of detector $A$ with the field and other detectors integrated out reads
\begin{eqnarray}
  \rho^R[K^A, \Delta^A; x^0] &=& \exp \left[{i\over\hbar}\left(\langle\hat{Q}^{}_A\rangle K^A -
    \langle\hat{P}^{}_A\rangle \Delta^A\right) \right.\nonumber\\ && \left.
    -{1\over 2\hbar^2} \left( K^A {\cal Q}_{AA} K^A
    -2 \Delta^A {\cal R}_{AA} K^A + \Delta^A {\cal P}_{AA} \Delta^A \right) \right].
\label{RDSQa}
\end{eqnarray}

Thus, looking at the evolution of the Gaussian state (\ref{Qstate}) or (\ref{RDSQa}) is equivalent to
looking at the dynamics of those symmetrized two-point correlators, which would be obtained
more easily in the Heisenberg picture.

\subsection{Correlator dynamics for factorizable initial states}

For mathematical convenience (and to a large extent reflective of not uncommon physical situations) one
often assumes that the initial state at $x^0_{\bf 0}=t_0$ in the Minkowski
frame is a product state of the Minkowski vacuum of the field 
(which is Gaussian) and the Gaussian state of the detectors $A, B, \cdots$. 
The detector part can be a product of the ground states and/or single-mode squeezed states,
a multi-mode squeezed state, or any mixed state in the Gaussian form.
The field part can be easily generated to a thermal state, which is Gaussian, too.

By virtue of linearity in (\ref{Stot1}), the operators of the detectors and the field in
the Heisenberg picture will evolve to a linear combination of all the detector
operators $\hat{Q}_{\bf d}$, $\hat{P}_{\bf d}$ 
and the field operators $\hat{\Phi}_{\bf k}$, $\hat{\Pi}_{\bf k}$ defined at the initial moment $t_0$.
Then each symmetrized two-point correlator of the detectors for the factorizable initial state
$\rho^{}_{\Phi_{\bf x}}\otimes\rho^{}_{\bf d}$ 
will split into a sum of the a-part and the v-part~\cite{LH06}. 
The a-part corresponds to the initial state of the detectors,
while the v-part corresponds to the response to the field vacuum $\left| 0^{}_M \right>$.

\subsection {Divergences}
\label{2reg}

There are two sources of the divergences for the correlators in this detector-field model.

\noindent{\bf 1.} 
In this linear system 
the mode functions  
satisfy the classical equations of motion (the only difference is 
the initial conditions). 
Thus they suffer the same divergences as the classical ones: 
In the UD detector theory in (3+1)D, the retarded field sourced by a pointlike detector diverges right at
the position of the detector. 
One needs to introduce a cutoff $\Lambda$ to regularize the $\delta$-function in the interaction
Hamiltonian between the detector and the field, expand the relevant mode functions of the detectors 
in series of $\Lambda$, absorb the divergent terms by some parameter of the model
(in our model it is the natural frequency of the detectors $\Omega$ \cite{LH06},
in other model it could be the mass of the detector $m$ or the coupling constant $\lambda$),
then take the $\Lambda\to \infty$ limit to eliminate the $O(\Lambda^{-1})$ terms. The $O(\Lambda^0)$ terms will
survive after taking the limit; it gives the radiation reaction such as the dissipation term ($\sim \gamma \dot{Q}$)
as those in our model or the higher derivative term 
as the one in the Abraham-Lorentz-Dirac equation \cite{Ro65}
\footnote{Note that there is no such divergence in the RSG model in (1+1)D \cite{RSG91}, where the retarded field
is regular everywhere and simply offers the radiation reaction in the equation of motion of the detector.}.

\noindent{\bf 2.} The second kind of divergences are the UV or IR divergences arised in the mode sum,
or equivalently, the divergences arised in the coincidence limit of the Green's functions (UV),
or in the integration over the whole position space (IR).
The regularization of the UV divergences should be consistent with those for the first kind.
This kind of divergences is also a common feature for 
quantum Brownian motion.


\subsection{Range of validity of perturbative results}

For a single Unruh-DeWitt (UD) detector moving in (3+1) dimensional Minkowski space,
The total action is given by (\ref{Stot1}) with ${\bf d}=A$.
Let us denote $Q\equiv Q_A$.
Suppose the initial state of the system at $\tau_0$ is a direct product of the ground state for $Q$
and the Minkowski vacuum for $\Phi$. 
It is straightforward to write down the reduced state of the detector in
the $(K,\Delta)$-representation. However, to compare with the perturbative results,
we look at the reduced density matrix $\rho^R (Q,Q';\tau)$ directly.
Transformed to the representation in the basis of energy eigenstates for the free harmonic oscillator $Q$, 
the transition probability from the initial ground state to the first
excited state then reads 
\begin{equation}
  \rho^R_{1,1} = { \hbar \left[\langle P^2 \rangle \langle Q^2 \rangle
  -\langle P,Q \rangle^2-(\hbar^2/4)\right]
  \over \left\{ \left[ \langle P^2 \rangle+ (\hbar m_0\Omega_r/2)\right]
  \left[\langle Q^2 \rangle+ (\hbar/ 2m_0\Omega_r) \right]-
    \langle  P, Q \rangle^2\right\}^{3/2}}
\label{rho11gen}
\end{equation}
where $\Omega_r$ is the renormalized frequency \cite{LH06}.
Expanding the symmetrized two-point correlators of the detector
in terms of the coupling strength $\gamma\equiv\lambda^2/(8\pi m_0)$,
the approximate value up to the first order of $\gamma$ becomes
\begin{equation}
  \rho^R_{1,1}|_{\gamma\eta\to 0} \stackrel{\eta\gg a^{-1}}{\longrightarrow}
  {\lambda^2\over 4\pi m_0}\left[ {\eta\over e^{2\pi\Omega_r/a}-1}
  + {\Lambda_1 +\Lambda_0-2\ln (a/\Omega_r) \over 2\pi \Omega_r} \right]
\label{rhopert}
\end{equation}
when $\eta\equiv \tau-\tau_0 \gg a^{-1}$. Here $\Lambda_0$ and $\Lambda_1$ are large constants
introduced by regularization (the second kind in Section \ref{2reg}).
We see that the first term of (\ref{rhopert}) gives the conventional transition probability 
from TDPT over infinite time. Only when $\Omega_r\eta \gg \Lambda_1, \Lambda_0$, or $a$ is extremely
large, can the second term in (\ref{rhopert}) be neglected. 
Hence the conventional transition probability or transition rate 
is valid only in the limits of (a) ultra-high acceleration ($a\gg \Omega_r$ and $\Lambda_1 \ll a\eta \ll
a\gamma^{-1}$) or (b) ultra-weak coupling  ($a^{-1},\Omega_r^{-1}\Lambda_1 \ll \eta \ll \gamma^{-1}$).
Only in these limits 
the effective temperature obtained by diagonalizing $\rho_{m,n}^R$ is very close to the Unruh temperature 
$T_U= \hbar a/2\pi k_B$ \cite{LH07}, and the thermal bath is only slightly affected by the
back reaction from the detector to the field, namely, the detector
acts essentially as a test particle in the field.

\begin{figure}
\includegraphics[width=5cm]{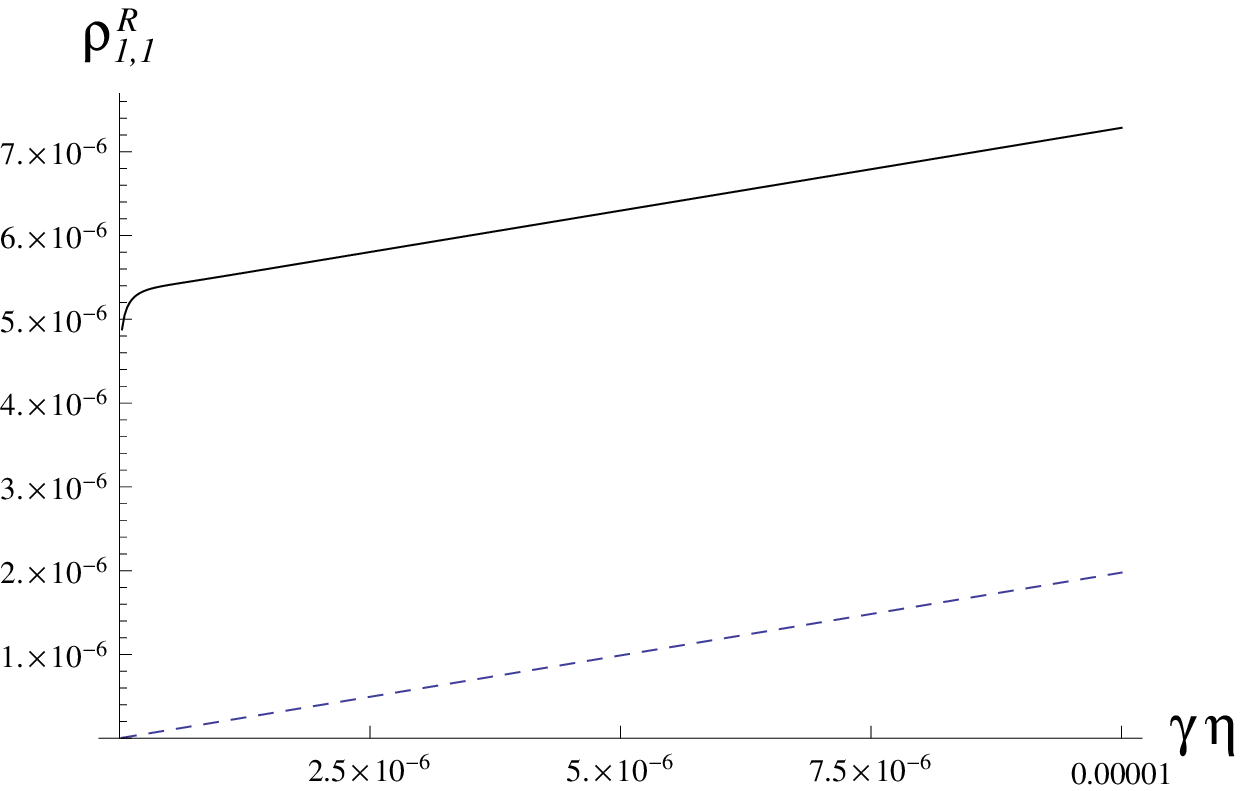}
\includegraphics[width=5cm]{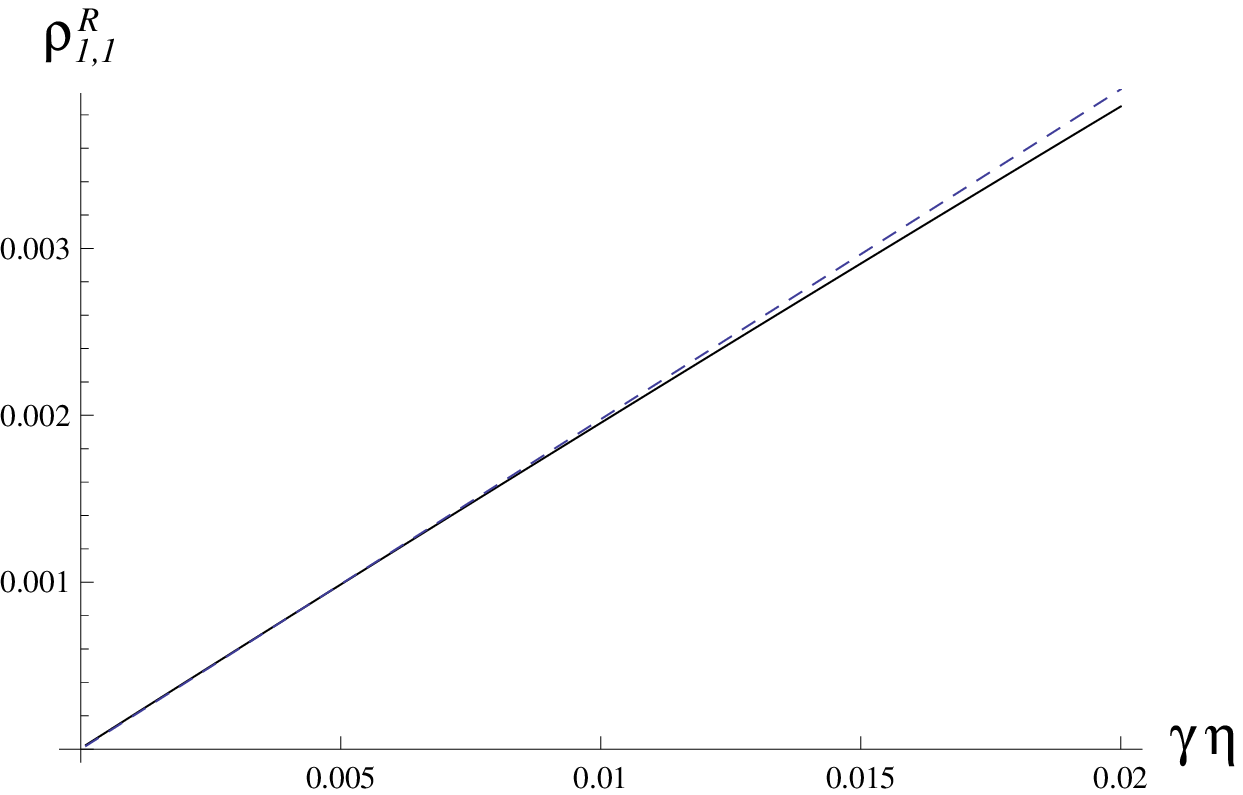}
\includegraphics[width=5cm]{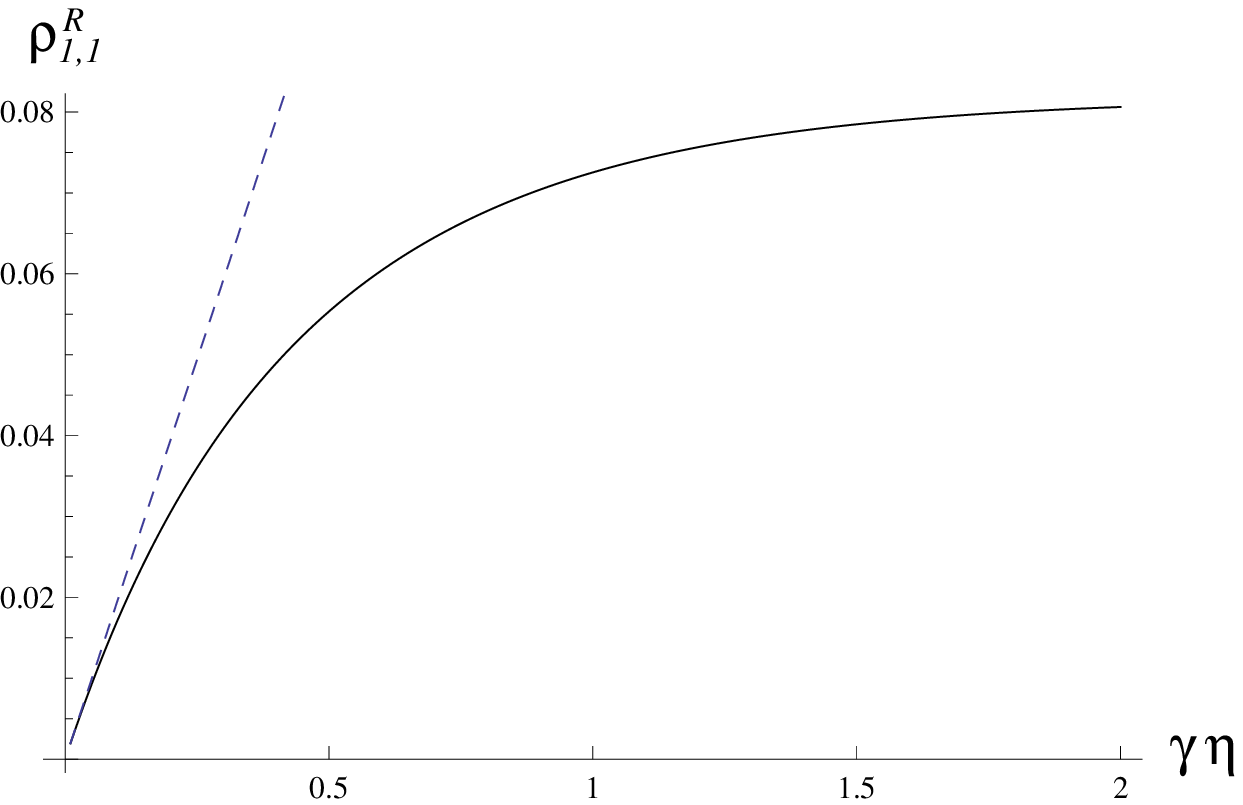}
\caption{Comparison of the exact $\rho^R_{1,1}$ (solid curve, Eq.(\ref{rho11gen})) and the perturbative
result (dashed line, the first term in (\ref{rhopert})) in different time scales of $\eta \equiv \tau-\tau_0$
for a uniformly accelerated detector initially in its ground state and coupled to the Minkowski vacuum of
the massless scalar field after $\tau_0$.
The parameters in these plots are $\gamma=10^{-6}$, $a=6$,
$\Omega = \sqrt{\Omega_r^2-\gamma^2}=2.3$, $\Lambda_0=\Lambda_1=20$, $m_0=1$, and $\hbar=c=1$.}
\label{rho11Evo}
\end{figure}

Note that, in obtaining (\ref{rhopert}), we have assumed $a^{-1} \ll
\eta \ll \gamma^{-1}$, when the system is still in transient. Indeed,
in figure \ref{rho11Evo} TDPT works well only in the middle plot 
(with $O(10^{-1}) < \eta < O(10^{-2}/\gamma)$ for $\gamma=10^{-6}$ and $a=6$).
If $a <\gamma$, the conventional transition probability 
has no chance to dominate at all. In particular, the $a=0$ case is beyond the reach of TDPT over infinite time,
and the conventional wisdom from perturbation theory that no transition occurs in an inertial detector
is untenable. In contrast, 
the evolution of $\rho^R_{1,1}$ with $a=0$ behaves qualitatively similar to those
cases with nonzero acceleration \cite{LH06}. This agrees with the
expectation from the observation that the UD detector theory is a
special case of quantum Brownian motion \cite{UZ89,HPZ,HM}.

$\Lambda_0$ corresponds to the time scale of switching on the
interaction, namely, the ``$\ln\delta$" in $d=4$ case in Section \ref{subsec:sharp}, so
it could be finite in real processes.
This implies that the $\Lambda_0$ terms in all two-point functions will be damped out so 
that $\Lambda_0$ will not be present in the late-time results.
Actually (\ref{rhopert}) is formally identical to the
first-order transition probability from TDPT for a UAD with
finite duration of interaction and $\Lambda_0$ and $\Lambda_1$ are
formally the same as the divergences found in \cite{Svaiter:1992xt}.
In \cite{Higuchi:1993cya} it has been shown that these divergences can be
tamed if one switches on and off the interaction smoothly, so can
$\Lambda_0$. Nevertheless, here we are looking at the real-time causal evolution problem
(``in-in" formulation) rather than a scattering transition amplitude
(``in-out" formulation) problem, and we never turn off the coupling, 
so $\Lambda_1$ is a non-zero constant of time.

$\Lambda_1$ should not be absorbed by any parameter or subtracted from any physical quantity of this theory 
for more reasons:
(a) The UD detector theory is not a fundamental theory to meet the
renormalizability requirement, and the presence of cut-offs as
physical parameters is an expected feature which characterizes the
range of validity of this effective theory, just like the
Compton wavelength of the electron serving as a cut-off in quantum optics;
(b) $\Lambda_1$ is not present in the renormalized stress-energy tensor of the field induced by the detector \cite{LH06},
so that $\Lambda_1$ is not observable outside of the detector;
(c) If $\Lambda_1$ was subtracted naively, the uncertainty principle will be
violated 
at late times for $a$ is small enough \cite{LH07}.

\section{Applications to RQI: quantum teleportation}

Quantum teleportation is not only of practical values but also of theoretical interest because it contains many
illuminating manifestations of quantum physics, clarifying fundamental issues such as quantum information and classical
information, quantum nonlocality and relativistic locality, spacelike correlations
and causality, etc. \cite{NielsenChuang, QTelepExpt1, QTelepExpt2} 

The first protocol of quantum teleportation is given by Bennett {\it et al.} 
in \cite{BBCJPW93}, where an {\it unknown} state of a qubit $C$ is teleported from one spatially localized agent Alice to
another agent Bob using an entangled pair of qubits $A$ and $B$ prepared in one of the Bell states and shared by Alice and Rob,
respectively. 
This idea is then adapted to the systems with continuous variables by Vaidman \cite{Va94}, 
who introduces an EPR state \cite{EPR35} for the shared entangled pair to teleport an unknown coherent state.
Braunstein and Kimble (BK) \cite{BK98} generalized Vaidman's scheme from EPR states with exact correlations
to squeezed coherent states. In doing so the uncertainty of the measureable quantities
reduces the degree of entanglement of the $AB$-pair as well as the fidelity of teleportation.

To explore how the Unruh effect affects teleportation,
Alsing and Milburn made the first attempt of calculating the fidelity of quantum teleportation between two moving cavities
in relativistic motions \cite{AM03} -- one is at rest (Alice), the other is uniformly accelerated (Rob) in the Minkowski frame
--  though their result is not quite reliable 
\cite{SU05, FM05, AF12}. Then Landulfo and Matsas \cite{LM09} 
considered quantum teleportation in the future asymtotic region in a two-level detector qubit model where Rob's detector 
is uniformly accelerated and interacting with the quantum field only in a finite duration. 
Alternatively, Shiokawa \cite{Tom09} has considered quantum teleportation in the UD detector theory 
with the agents in similar motions but based on the BK scheme in the interaction region.
More recently, the relativistic effects of quantum information associated with the quantum field have been taken into
account carefully in \cite{LSCH12}, as summarized below.

Consider a setup with detectors $A$ and $C$ held by Alice, who is at rest in space with the
worldline $\z_A^{} =\z_C^{} = (t, 1/b,0,0)$ in the Minkowski frame, 
and detector $B$ held by Rob, who is uniformly accelerated along the worldline
$\z_B^{} = (a^{-1}\sinh a\tau, a^{-1}\cosh a\tau,0,0)$, $0<a<b$, where $\tau$ is Rob's proper time,
namely, $\tau^{}_A=\tau^{}_C=t$ and $\tau^{}_B=\tau$.
Suppose the initial state of the combined system at $t=\tau=0$ is a product state $\rho^{}_{\Phi_{\bf x}}\otimes\rho^{}_{AB}
\otimes\rho_C^{(\alpha)}$ of the Minkowski vacuum of the field $\hat{\rho}^{}_{\Phi_{\bf x}} = \left| 0_M\right>\left< 0_M\right|$,
a two-mode squeezed state $\rho^{}_{AB}$ of the detectors $A$ and $B$,
and a coherent state of the detector $C$, denoted $\hat{\rho}_C^{(\alpha)} =\left|\alpha \right>^{}_C\left<\alpha\right|$,
which is the quantum state to be teleported.
To concentrate on the best fidelity of quantum teleportation that the entangled $AB$-pair can offer, however,
we assume the dynamics of $\rho_C^{(\alpha)}$ is frozen.
Also we design $\rho^{}_{AB}$ so that it goes to an EPR state with the correlations
while $Q_A+Q_B$ and $P_A-P_B$ are totally uncertain as its squeezed parameter $r_1\to \infty$.

In the Minkowski frame, at $t=0$, the detectors $A$ and $B$ start to couple with the field,
while the detector $C$ is isolated from others. At $t=t_1$ when the reduced state 
of the three detectors continuously evolves to $\rho^{}_{ABC}({\bf K},{\bf \Delta};t_1)$, 
a joint Gaussian measurement by Alice is performed {\it locally in space} on $A$ and $C$ so that the post-measurement 
state right after $t_1$ collapses to $\tilde{\rho}^{}_{ABC}({\bf K},{\bf \Delta};t_1)= 
\tilde{\rho}^{(\beta)}_{AC}(K^A, K^C,\Delta^A, \Delta^C)\tilde{\rho}_{B}(K^B, \Delta^B)$,
where $\tilde{\rho}^{(\beta)}_{AC}(K^A, K^C,\Delta^A, \Delta^C)$ is a two-mode squeezed state of
detectors $A$ and $C$ with displacement $\beta = \beta_R+i\beta_I$, which is the outcome Alice obtains.

\subsection{Entanglement and Pseudo-Fidelities in Different Frames}
\label{EnPsFi}

Quantum entanglement of the detector pair $A$ and $B$ in a Gaussian state is fully determined by the 
symmetrized two-point correlators
${\cal Q}_{ij}$, ${\cal P}_{ij}$, and ${\cal R}_{ij}$, $i,j=A,B$ in the quantum state
(\ref{Qstate}) \cite{LCH08,Si00,VW02}.
Since the worldlines of the two detectors do not intersect, the
entanglement between the detectors at some moment will be a kind of spacelike correlation
and depend on the spatial hypersurface on which the entanglement is defined.  
The time evolution of the entanglement will hence depend on how the spacetime is foliated
by spacelike hypersurfaces.
In general entanglement of the detectors is incommensurable with the physical fidelity of quantum teleportation,
which is a kind of timelike correlation.

\begin{figure}
\includegraphics[width=4.5cm]{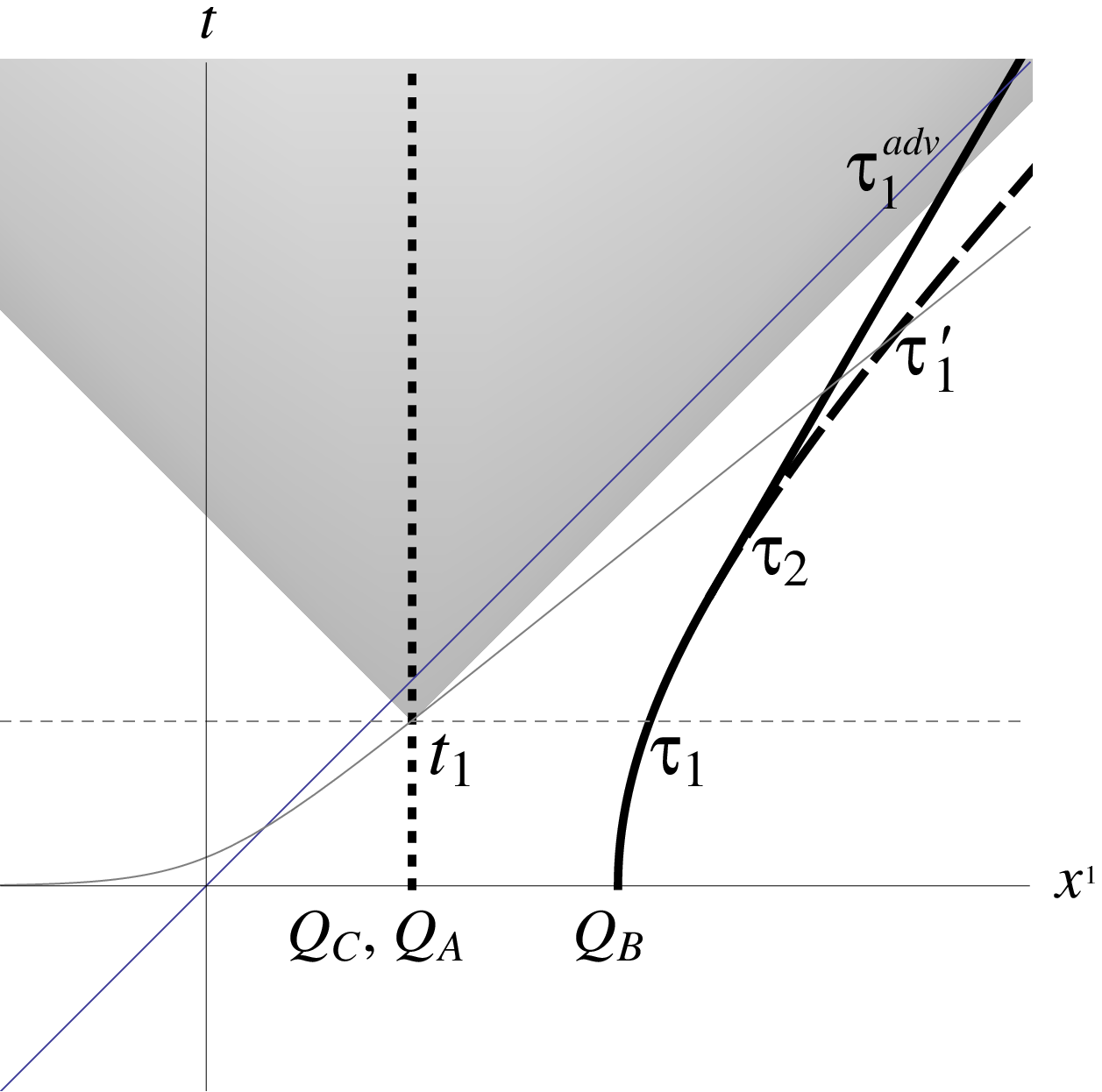}\hspace{.5cm}
\includegraphics[width=5cm]{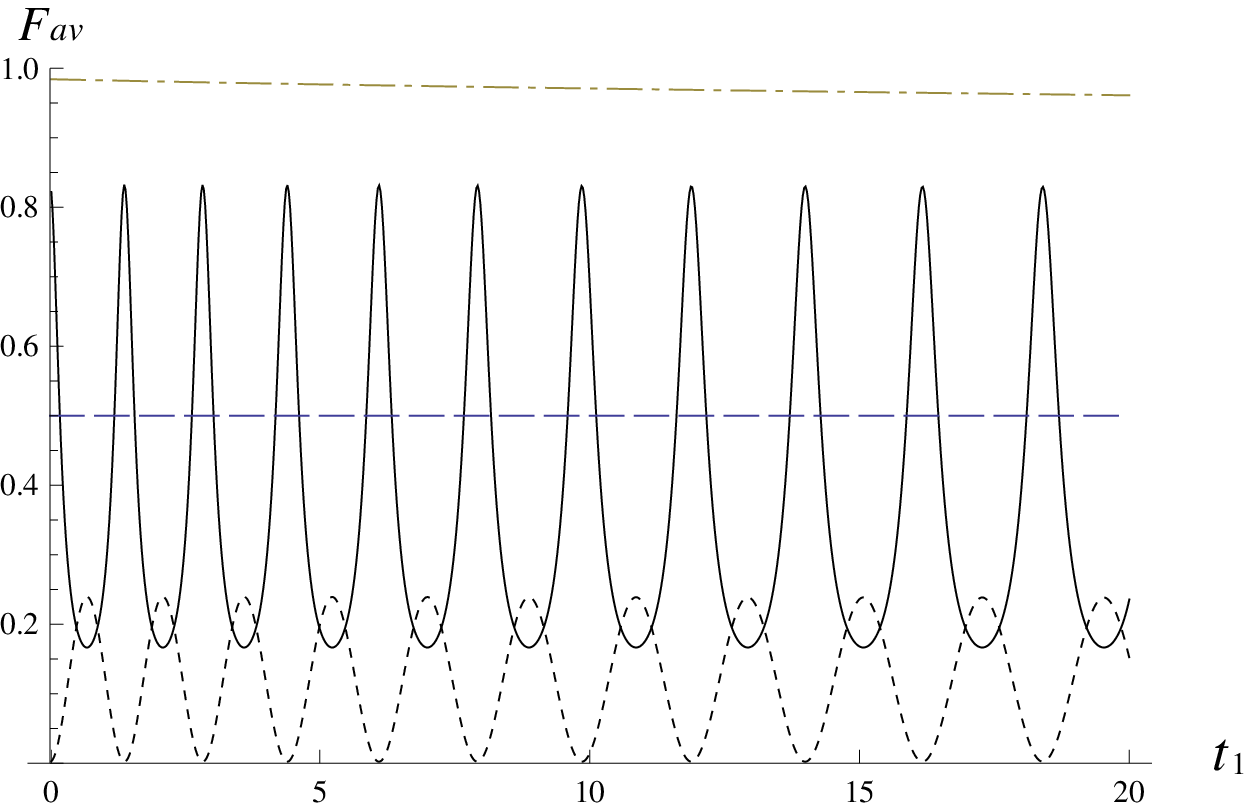}
\includegraphics[width=5cm]{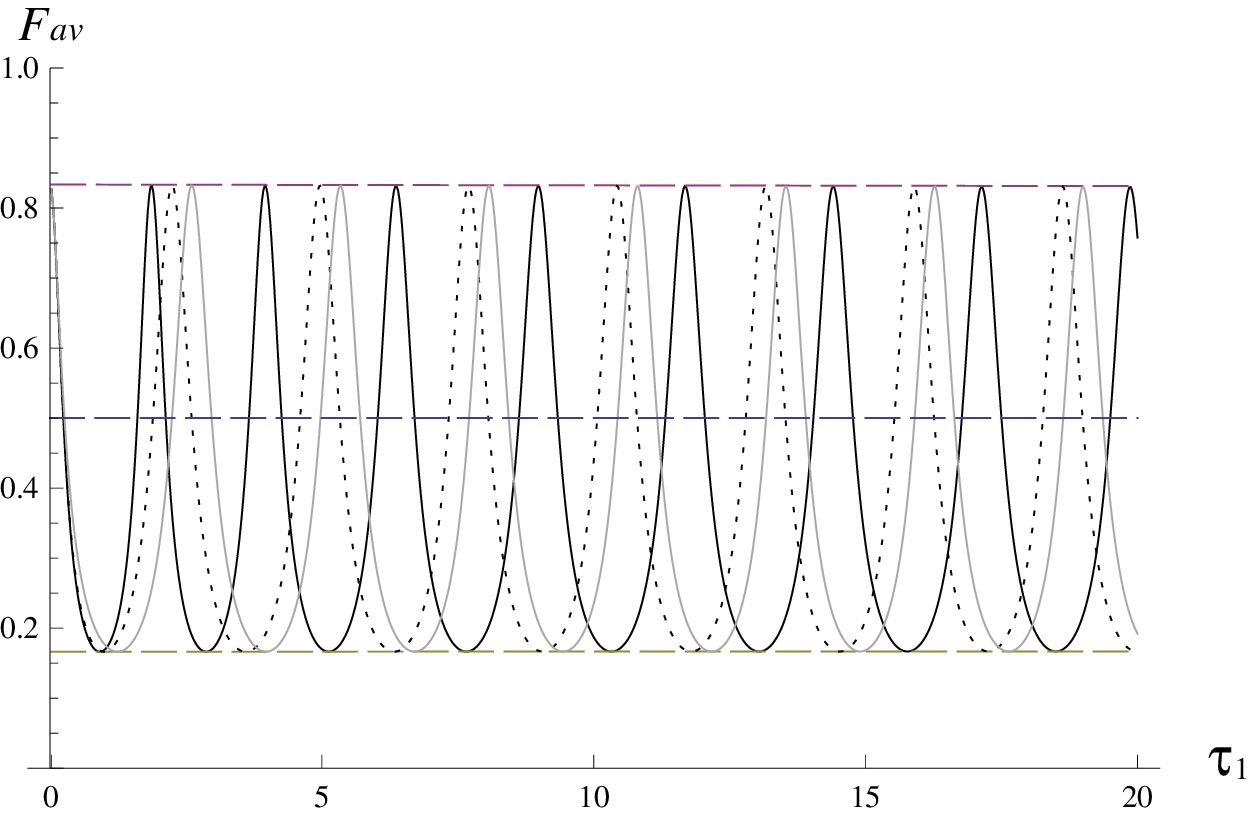}
\caption{(Left) Setup for quantum teleportation from Alice (thick dotted line)
to Rob (thick dashed for section \ref{EnPsFi} and thick solid for \ref{physreal}).
The gray solid curve and the horizontal line represent a $\tau_1'$-slice in the quasi-Rindler frame
and a $t_1$-slice in the Minkowski frame, respectively.
The shaded region is in the future lightcone of the measurement event 
by Alice, and the hypersurface $t=x^1$ is the event horizon of Rob for $\tau_2\to\infty$.
(Middle) A comparison of the averaged pseudo-fidelity $F_{av}$ (solid) in the Minkowski frame, 
the correlator $\langle Q_-^2 \rangle/20$ (dotted), $Q_-\equiv Q_A - Q_B$
and the logarithmic negativity $E_{\cal N}/3.5$ (dot-dashed) at early times  in the ultraweak coupling limit.
(Right) $F_{av}$ in the quasi-Rindler frame with the same parameters (solid) and with different $a$ (dotted, gray).
The upper and middle dashed lines represent $F^+_{av}$ and $1/2$.}
\label{AR}
\end{figure}

To compare these two correlations,
let us first imagine that Rob receives the outcome $\beta$ of Alice's joint measurement 
and make the proper operation on detector $B$ {\it instantaneously} at $\tau_1(t_1)$ when the worldline of $B$
intersects the $t_1$-slice (see figure \ref{AR}).
According to the outcome $\beta$ obtained by Alice,
the operation that Rob should perform on detector $B$ is a displacement by $\beta$ in phase space of $B$
from $\tilde{\rho}^{}_B$ to $\rho^{}_{out}$, which is defined on a time-slice right after the
one where the post-measurement state $\tilde{\rho}^{}_{ABC}$ is defined.
The ``pseudo-fidelity" of quantum teleportation from $|\alpha\rangle^{}_C$ to
$|\alpha\rangle^{}_B$ is defined as
$F(\beta) \equiv {}^{}_B\langle \alpha\,|\hat{\rho}_{out}|\,\alpha\rangle^{}_B / {\rm Tr}_B\hat{\rho}^{}_{out}$,
where ${\rm Tr}^{}_B\hat{\rho}^{}_{out} = 
P(\beta)$ has been normalized to be the probability of finding the outcome $\beta$.
Then the {\it averaged} pseudo-fidelity is defined by
\begin{equation}
  F_{av} \equiv \int d^2 \beta P(\beta) F(\beta) = \int d\beta_R d\beta_I \,
   {}^{}_B\langle\alpha|\hat{\rho}_{out}|\alpha\rangle^{}_B . 
\label{Favformula}
\end{equation}
$F_{av}=1/2$ is known as the best fidelity of ``classical" teleportation using coherent states \cite{BK98},
without considering the coupling of the UD detectors with the environment.

Two results in the ultraweak coupling limit, one in the Minkowski frame, the other in the ``quasi-Rindler frame"
\footnote{By a quasi-Rindler frame we refer to the coordinate system in which each time-slice almost overlaps a Rindler time-slice in
the R-wedge but the part in the L-wedge has been bent to the region with positive $t$ to make the whole time-slice located
after the initial time-slice for the Minkowski observer, as illustrated in figure \ref{AR}.}
are shown in figure \ref{AR} (middle, right).
One can see that the degree of entanglement (logarithmic negativity $E_{\cal N}$) evolves smoothly while the averaged
pseudo-fidelity evaluated in whatever reference frame oscillates in $t_1$ or $\tau_1$.
Even at very early times $F_{av}$ drops below $1/2$ frequently when $E_{\cal N}$ is still large. 
Clearly the oscillation of $F_{av}$ 
here is mainly due to the distortion of the quantum state of $AB$-pair 
from their initial state (caused by the alternating squeeze-antisqueeze natural oscillations of their quantum state)
rather than the disentanglement between them.

In our setup, when $t_1$ gets larger, the time dilation of detector $B$ becomes more significant
and so detector $B$ appears to change extremely slowly in the Minkowski frame, while in the quasi-Rindler frame,
it is the time-dilation of detector $A$ become obvious and so
detector $A$ looks frozen for the Rindler observer when $\tau_1$ gets larger.
In both cases 
it makes the frequency of the oscillation of $F_{av}$ approaches $\Omega$ for larger times.

The peak values of $F_{av}$, denoted by $F^+_{av}$, fall below the fidelity of classical teleportation $1/2$
(at the moment denoted $t_{1/2}$ or $\tau_{1/2}$) always earlier than the dis-entanglement time $t_{dE}$ or $\tau_{dE}$ when 
$E_{\cal N}$ become zero in both frames. Indeed, in Appendix A of \cite{LSCH12} it is shown that 
entanglement between detectors $A$ and $B$ is necessary to provide the advantage of quantum teleportation,
at least in the ultraweak coupling limit. 

In the Minkowski frame, both $F^+_{av}$ and $E_{\cal N}$ are insensitive to $a$ in the ultraweak coupling limit.
In contrast, the dependence on the proper acceleration $a$ is obvious in the quasi-Rindler frame, where
the larger $a$, the earlier $\tau_{1/2}$ and the earlier disentanglement time $\tau_{dE}$.
Here we see the frame dependence of entanglement dynamics.

Beyond the ultraweak coupling limit, both $F_{av}$ and $E_{\cal N}$ are strongly affected by the environment.
In most cases quantum entanglement disappears quickly both in the Minkowski frame and the quasi-Rindler frame
due to strong interplays with the environment,
and the averaged pseudo-fidelity $F_{av}$ drops below $1/2$ even quicker.

\subsection{Physical fidelity and ``entanglement on the lightcone"}
\label{physreal}

Suppose Rob stops accelerating at his proper time $\tau_2$,
after this moment Rob moves with constant velocity,
while Alice stays at rest 
and performs the joint measurement on A and C at $t_1$ (see figure \ref{AR}).
In this setup the classical information from Alice traveling at the speed of light can always reach Rob,
though the acceleration of detector $B$ is not uniform -- For the dynamics of the correlators
similar situations we refer to Ref. \cite{OLMH11}, more results can be found in \cite{LinDICE10}.

Suppose Rob performs the local operation on $B$ at some moment $\tau^{}_P > \tau^{adv}_1$ when he received
the information traveling in lightspeed from Alice (see figure \ref{AR} for the definition of $\tau^{adv}_1$).
Then, similar to (\ref{Favformula}), the averaged physical fidelity should be given by
$F_{av} = \int d^2\beta \,\,{}^{}_B\hspace{-.07cm}\left<\alpha\right| \hat{\rho}^{}_{out}(\tau^{}_P)\left|\alpha\right>^{}_B$,
where $\rho^{}_{out}(\tau^{}_P)$ is obtained by performing a displacement on $\tilde{\rho}^{}_B(\tau^{}_P)$ which started with
the post-measurement state $\tilde{\rho}^{}_B(\tau_1)$ with $\tau_1 = a^{-1}\sinh^{-1}a t_1$ (when the quantum state collapses
in the Minkowski frame) and evolves from $\tau_1$ to $\tau^{}_P$. 
Nevertheless, an analysis similar to \cite{Lin11a} shows that
the correlators in the reduced state of detector $B$ observed in all reference frames
will become the same collapsed ones 
at the moment when detector $B$ is entering the future lightcone of
the measurement event by Alice ($\tau^{}_B = \tau^{adv}_1$). 
So we are allowed to collapse the wave functional on a time-slice intersecting Alice's worldline 
at $\tau^{}_A = t_1$ and Rob's at $\tau^{}_B = \tau^{adv}_1-\epsilon$, $\epsilon \to 0+$.

If we further assume that mutual influences are small and Rob performs the local operation at 
$\tau^{}_P = \tau^{adv}_1+\epsilon$ right after the classical
information from Alice is received, 
then the continuous evolution of the reduced state of detector $B$ from $\tau^{}_M$ to $\tau^{}_P$ is negligible,
and so we can directly compare the physical fidelity at $\tau^{adv}_1+\epsilon$ with entanglement between detectors $A$ at $t_1$ 
and $B$ at $\tau^{adv}_1-\epsilon$. Again from 
Appendix A of \cite{LSCH12} with the proper time of detector $B$ substituted by $\tau_1^{adv}$
(actually $\tau_1^{adv}\pm \epsilon$, $\epsilon\to 0+$),  
quantum entanglement of $AB$-pair evaluated almost on the future lightcone of the measurement event
by Alice is still a necessary condition of the best averaged physical fidelity of quantum teleportation beating the
classical one in the ultraweak coupling limit.

\begin{figure}
\includegraphics[width=5cm]{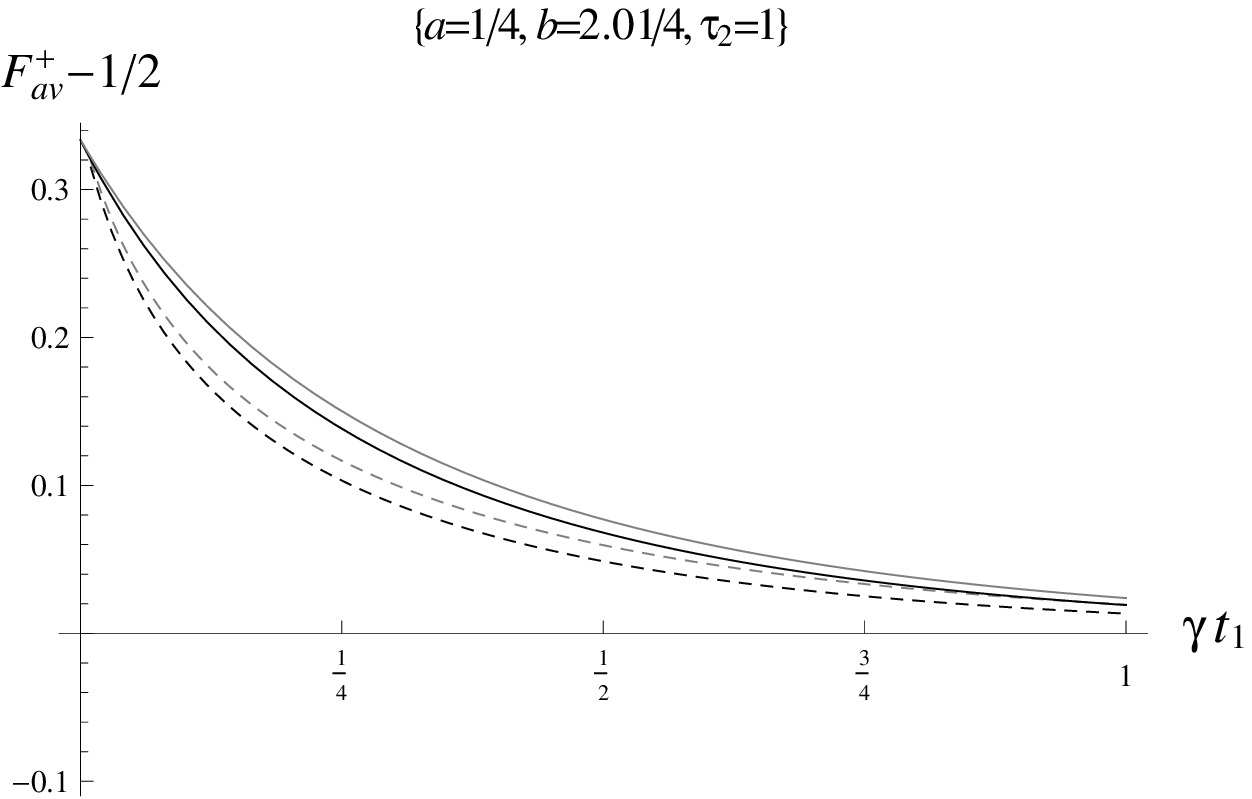}
\includegraphics[width=5cm]{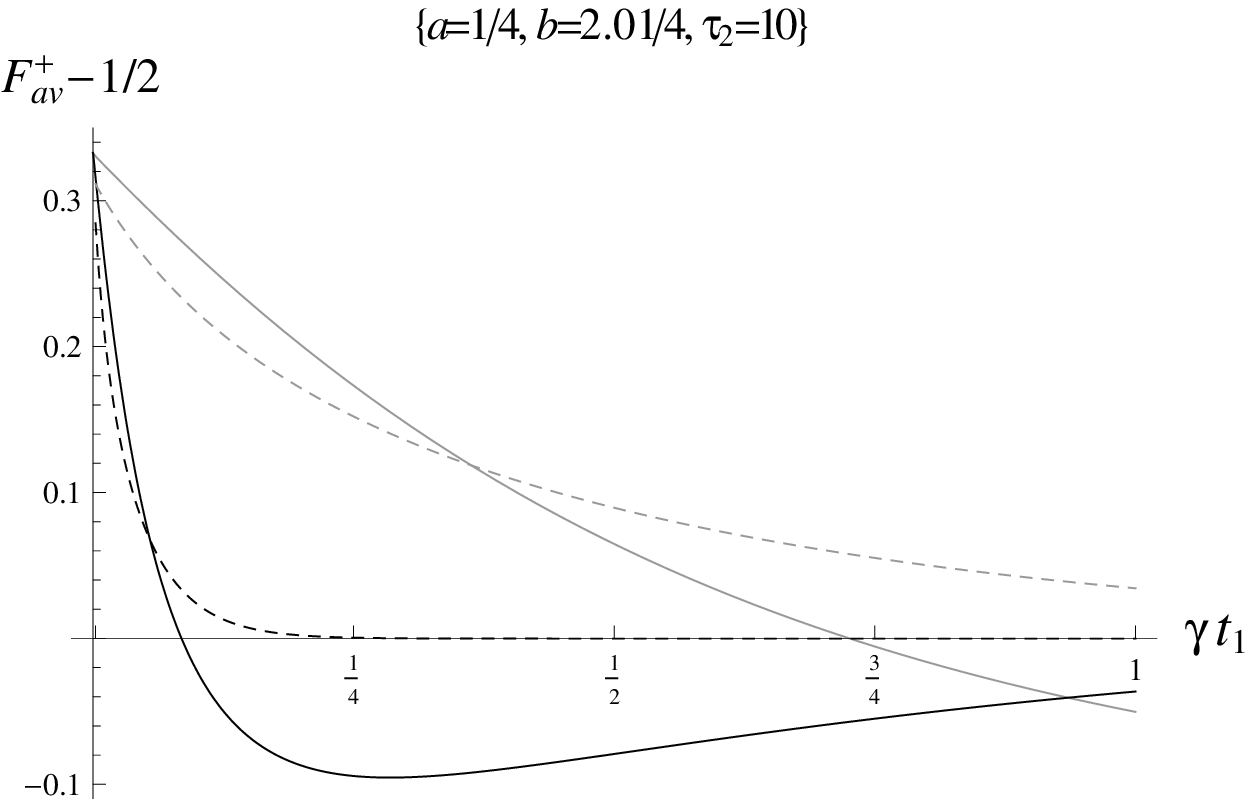}
\includegraphics[width=5cm]{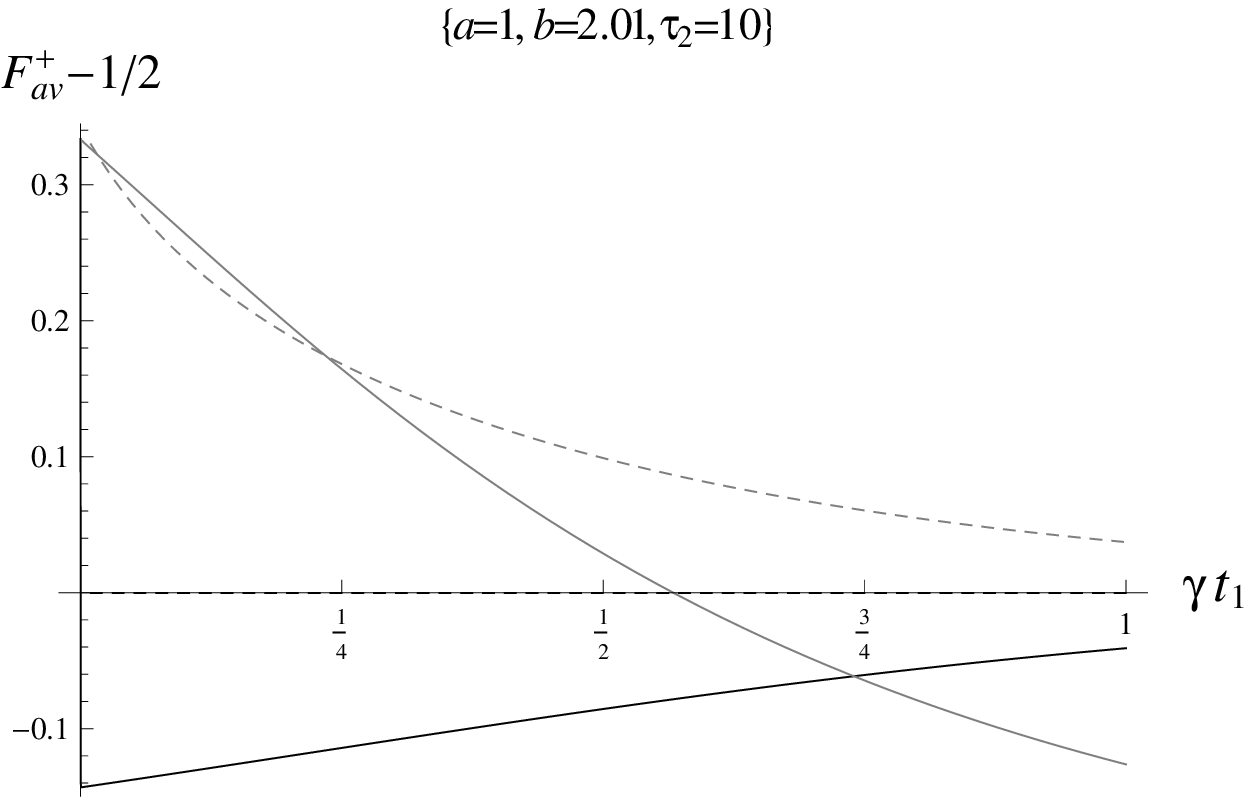}
\caption{$F_{av}^+-1/2$ (solid) and $E_{\cal N}/10$ (dashed) in section \ref{physreal} 
as a function of the moment of the joint measurement $t_1$.
The black curves are those physical fidelities and the degrees of entanglement evaluated ``on the lightcone",
while the gray curves are those pseudo-fidelities and the degrees of entanglement evaluated
on the $t_1$-slice in the Minkowski frame.}
\label{FavPlus}
\end{figure}

The number of peaks of the physical $F_{av}$ in the same duration of $t_1$ in this more
realistic case is much more than the one for the averaged pseudo-fidelity, because it takes a long time from 
$\tau_1(t_1)$ to the moment $\tau_1^{adv}$ when the classical signal from Alice reaches Rob,
during which detector $B$ has oscillated for many times.
In figure \ref{FavPlus} we see that the moment $t_1=t_{1/2}$ when the best averaged physical fidelity of quantum
teleportation $F_{av}^+$ drops to $1/2$ is earlier than any $F_{av}^+$ of pseudo-fidelity has.
The larger $a\tau_2$, the later $\tau^{adv}_1$ Rob has, 
and so the lower value of the physical $F_{av}^+$ at that time due to the longer time
of coupling with environment. When $a\tau_2$ is large enough, $\tau_1^{adv}$ is so large that
$t_{1/2}$ is almost the moment that Alice enters the event horizon of Rob for $\tau_2\to\infty$.

\section{Outlook}

Detector-field interaction is a very useful and versatile system to expound and explore many known and unknown effects
of relativistic quantum information. Continuing the vein of quantum teleportation as example, it is not difficult to
generalize the systematics to curved spacetimes, from weak gravitational field as in the Earth's environment to
black hole spacetimes. 


In another vein, learning from the new techniques and ideas in the study of detector-field interaction as was done for the Unruh effect \cite{LH06,LH07}, and using the well-known correspondence between the Rindler and the Schwarzschild spacetimes, one may go beyond the test-field description of black hole physics and study how backreaction from the field impacts on the evolution of the black hole and the ``information loss" issues.\\



\noindent{\bf Acknowledgment}   This work is supported in part by
NSF under Grants PHY08-01368, 
PHY11-25915, the NSC Taiwan under Grant
99-2112-M-018-001-MY3, and the National Center for Theoretical Sciences, Taiwan.
J.~L. thanks Gabor Kunstatter for hospitality at the University of Winnipeg
and the organisers of the ``Bits, Branes, Black Holes'' programme for hospitality at the
Kavli Institute for Theoretical Physics, University of California at Santa Barbara.
J.~L. was supported in part by STFC (UK).
\\ \\

\end{document}